\documentclass[preprint]{elsarticle} 

\usepackage{amssymb}
\usepackage{amsmath}
\usepackage{algorithm}
\usepackage{algorithmic}
\usepackage{amsfonts}
\usepackage{microtype}

\title{Revisiting the Majority Problem: Average-Case Analysis with Arbitrarily Many Colours}
\author[mmu]{A. Kleerekoper\corref{cor1}}
\ead{a.kleerekoper@mmu.ac.uk}
	
\cortext[cor1]{Corresponding author}
\address[mmu]{SCMDT, Manchester Metropolitan University, Chester Street, Manchester, UK}

\begin{document}

\begin{abstract}
	The majority problem is a special case of the heavy hitters problem. Given a collection of coloured balls, the task is to identify the majority colour or state that no such colour exists. Whilst the special case of two-colours has been well studied, the average-case performance for arbitrarily many colours has not. In this paper, we present heuristic analysis of the average-case performance of three deterministic algorithms that appear in the literature. We empirically validate our analysis with large scale simulations.
\end{abstract}


\maketitle

\section{Introduction}

The majority problem is a special case of the heavy hitters problem. Indeed, research into the more general case was prompted by one of the solutions to the majority problem (see \cite{Boyer1991}, Sec. 5.8 and \cite{Misra1982}).  

Given a collection of $n$ balls, $\{x_1, ... x_n\}$, each of which is coloured with one of $m$ colours, the majority problem is to identify the colour that appears on more than $n/2$ of the balls, or to state that no such colour exists.

Boyer and Moore were the first to propose a solution which they called MJRTY \cite{Boyer1991}. Soon after, Fischer and Salzberg provided an algorithm with optimal worst-case performance \cite{Fischer1982}. Matula proposed an alternative algorithm with the same worst-case performance that we refer to as the Tournament algorithm \cite{Matula1990}. 

All three algorithms are based on pairing balls of different colours. When there is a majority, we are guaranteed that after such a process the only unpaired balls remaining will be of the majority colour. The analysis of this problem has therefore focused on the number of comparisons. Fischer and Salzberg proved that, when there are arbitrarily many colours, $\lceil\frac{3n}{2}\rceil-2$ comparisons are necessary and sufficient \cite{Fischer1982}. Matula repeated the proof independently \cite{Matula1990}

Aside from the proofs for the worst-case, all the analysis of the majority problem has been for the special case where there are only two possible colours. In that case, it has been shown that $n-\nu(n)$ comparisons are necessary and sufficient, where $\nu(n)$ is the number of 1s in the binary expansion of $n$ \cite{Saks1991,Alonso1993,Wiener2002}. Assuming that all $2^n$ possible inputs are equally probable, the average-case complexity has been shown to be lower-bounded by $\frac{2n}{3}-\sqrt{\frac{8n}{9\pi}}+\Theta(1)$ and that $\frac{2n}{3}-\sqrt{\frac{8n}{9\pi}}+O(\log n)$ comparisons are necessary and sufficient \cite{Alonso1997}. Under the same assumption, it has been shown that the average-case complexity of Boyer and Moore's MJRTY algorithm is $n - \sqrt{2n/\pi} +O(1)$ \cite{Alonso2013}.

In this paper, we consider the average-case complexity of the three deterministic algorithms. We assume that there are an arbitrary number of colours and that all $n^m$ possible streams are equally likely. Our analysis is heuristic but we empirically validate our analysis with large scale simulations.

\section{Boyer and Moore's MJRTY Algorithm}

\subsection{Brief Description of the Algorithm}

Boyer and Moore's algorithm, MJRTY, works by simulating a process of pairing balls of different colours and discarding the pairs. The algorithm uses two variables: One to store the candidate colour and one to store the number of unpaired balls of that colour. Whenever the counter is zero, the colour of the next ball drawn becomes the candidate colour; otherwise the ball that is drawn is compared to the candidate colour. If the new balls has the same colour as the candidate colour, then the counter is incremented because we have encountered another unpaired ball. If the new ball has a different colour, then the counter is decrement which represents the pairing and discarding of two unmatched balls.

When this process completes, the candidate colour is the colour of the majority of balls, if a majority exists. A second pass over the stream is required to verify whether the candidate colour appears on a majority of balls. This is done by simply counting the number of times that colour is seen. 

The pseudocode for the algorithm can be found in the Appendix.

\subsection{Analysis of the Algorithm}

The number of comparisons required in the second (verification) phase is always $n$ because we have to compare every ball to the candidate without exception. The number of comparisons required in the first phase is $n$ less the number of times the counter is zero, because that is the only situation in which a comparison is not performed.

The key to our analysis is to note that, although there may be an arbitrary number of colours, the problem reduces to a two-colour problem as soon as a candidate is selected. This is because the candidate counter is incremented for one colour and decremented for all other colours. When counting the number of comparisons, therefore, we only need to consider the number of comparisons needed in a two-colour problem.

Let $p$ be the probability of incrementing the counter and $q = 1-p$ be the probability of decrementing the counter. We can model the algorithm's behaviour as a random walk in which we walk \textit{up} with probability $p$ and \textit{down} with probability $q$. The number of times this walk hits the horizon is equivalent to the number of times the candidate counter returns to zero. 

To calculate the expected number of times the walk hits the horizon, we could consider Dyck paths, since a return to zero must start at zero and return to zero. We could then create an indicator variable, $X_i$ which is 1 if the walk is at zero after $2i$ steps and 0 otherwise. Finding the expected value of $X$ would give the expected number of times the counter is zero. 

However, in our case the horizon is reflective such that whenever the walk is at the horizon the next step is an \textit{up}, with probability 1. Therefore, we would need to consider the Catalan Triangle numbers, $T(i,k)$, which give the number of paths of length $2i$ having $k$ returns to zero \cite[A033184]{Sloane2007}. To find the expected number of times the walk hits the horizon, we would need to sum over all values of $i$ up to $n/2$ and for each value of $i$, another sum over all values of $k$ up to $i$. This has no closed form and is extremely expensive to calculate as $n$ increases.

An alternative, heuristic analysis is to consider the expected number of balls that are drawn between the counter being zero and returning to zero. In fact, we must find the expected number of balls that are drawn between the counter being one and returning to zero because whenever the counter is zero it must move to one on the next draw.

This can again be modelled as a asymmetric random walk. The question now becomes the expected number of steps to go from position 1 to 0. Faris provides the solution in section 1.2 of his work Lectures on Stochastic Processes \cite{Faris2001}. The solution is $1(q-p)$. In our case, we have $p = 1/m$ and $q = 1 - 1/m$, which makes the expected number of steps required to go from one to zero equal to $m/(m-2)$.

Let $t$ be the expected number of steps required to go from one to zero. $1+t$ will therefore be the expected number of balls drawn when the counter moves from zero back to zero again. Heuristically, we can say that the expected number of zeros is $n/(1+t) = n/(1+m/(m-2))$. The expected total number of comparisons is therefore:

\begin{equation}
E[C] = 2n - 1 - n/(1+\frac{m}{m-2})
\label{eqn:MJRTY}
\end{equation}

We note that this equation has no solution when there are two colours, but the average number of comparisons required in that case has already been provided by Alonso and Reingold \cite{Alonso2013}. 

\subsection{Experimental Validation}

We validate our analysis with large scale simulations by running the MJRTY algorithm on synthetic data. We vary the number of colours, selecting 3, 5 and 10 colours. For each colour, we vary the stream length between 250 and 10,000 in steps of 250. For each stream length we create 100 streams with colours represented as integers. The streams are created so that every ``colour'' is assigned to each item uniformly at random. 

For every stream, the number of comparisons used by the MJRTY algorithm is recorded and compared to the expected number, using equation (\ref{eqn:MJRTY}). The expected number of comparisons is found to always be within the 99\% confidence interval of the mean number of observed comparisons. Furthermore, the relative difference between the mean of the observed number of comparisons and the expected number is on average 0.05\%, 0.04\% and 0.03\% for 3, 5 and 10 colours respectively.

\begin{figure}[t]
\centering
\includegraphics[width=\columnwidth]{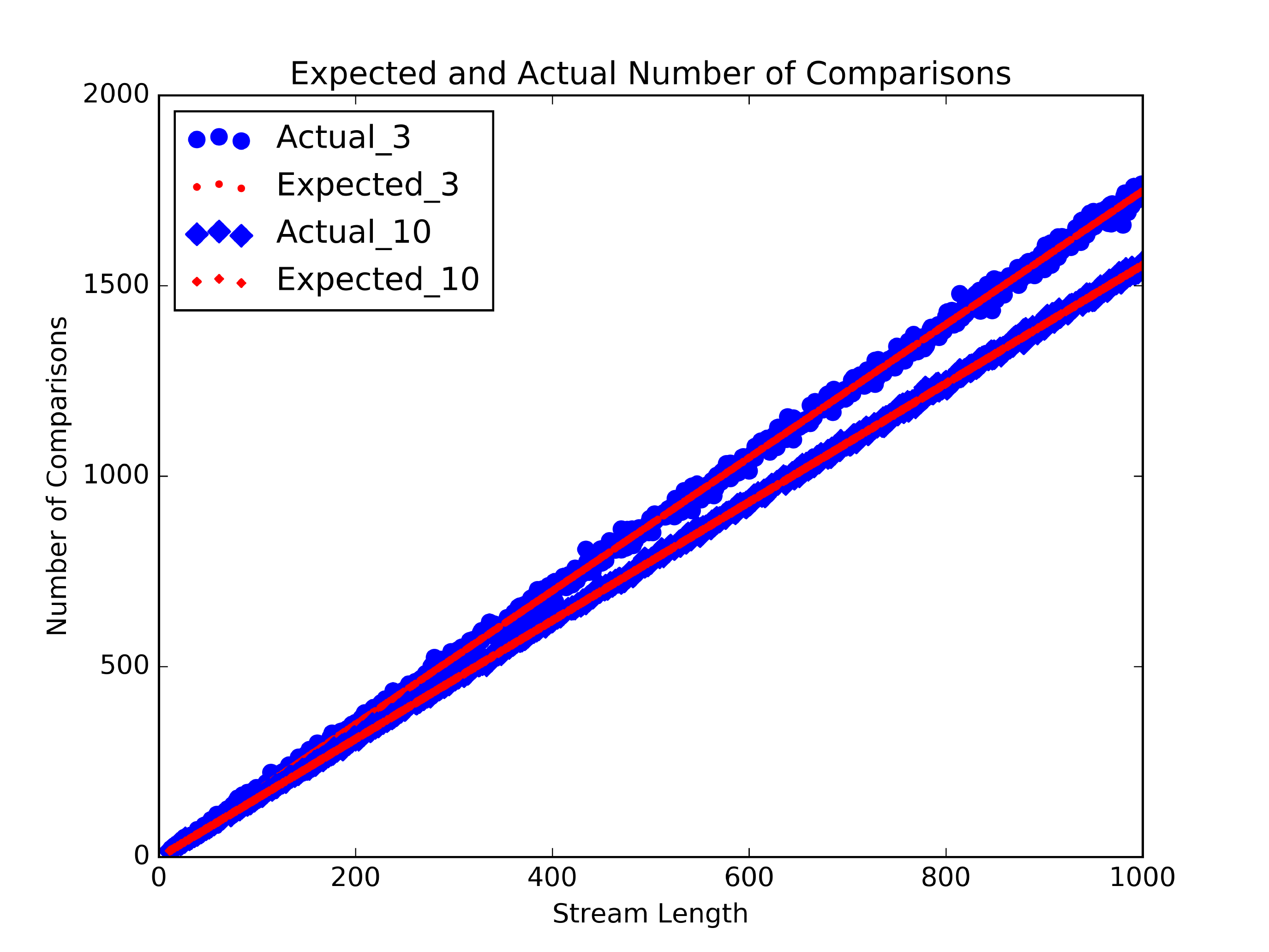}
\caption{A comparison of the expected and actual number of comparisons for the MJRTY algorithm for 3 and 10 colours - best viewed in colour.}
\label{fig:MJRTY}
\end{figure}

For a visual example we simulate 1,000 streams each of which is a random length between 10 and 1,000. For each stream we calculate the actual number of comparisons used and compare to the expected number. We repeat the experiment for 3 and 10 colours. Fig. \ref{fig:MJRTY} shows the results.

\section{Matula's Tournament Algorithm}

\subsection{Brief Description of the Algorithm}

The Tournament algorithm, first proposed by Matula, is also based on the pairing off of balls but it stores the results of those comparisons to speed up verification. The algorithm maintains a set of lists where each list has an associated weight ($w = 0,1,2,3\dots$) such that every item in the list is a shorthand for $2^w$ balls of the same colour. There is also a list of discarded tuples of the form $\{v',v'',j\}$ where $v'$ and $v''$ are balls of different colours both of which are shorthand for $2^j$ balls of that colour.

Initially all balls are in the list with weight 0 (this is just the initial stream). Then, in each round, pairs of balls from the same list are compared and either merged or discarded. If the two balls have the same colour, then they are merged and placed as a single entry in the next list. If they do not have the same colour, then they are formed into a tuple and added to the list of discarded tuples. The process continues until no list contains more than one ball. At this point, the candidate ball is the one in the most heavily weighted list which contains a ball.

A verification phase is needed in which the lists are traversed in reverse order and balls (really representing collections of balls) are compared to the candidate. A counter is maintained that is incremented by the appropriate amount for each list. Once all the lists are traversed, the discarded tuples must be examined as well.

The pseudocode for the algorithm can be found in the Appendix.

\subsection{Analysis of the Algorithm}

In the first round of the Tournament algorithm there are $n$ balls and we therefore need $n/2$ comparisons. A ball is entered into the next round if, upon comparison, the second ball matches the first which happens with probability $1/m$. So we would expected to find $n/2m$ balls in the second round. In the second round we perform $n/4m$ comparisons and expect $n/4m^2$ balls in the third round. In general, we expect to perform $n/2^im^{i-1}$ comparisons in round $i$, starting with round 1. There are a maximum of $\log_2(n)$ rounds but if we sum to infinity, the sum simplifies. The expected number of comparisons in the first phase is therefore:

\begin{equation}
E[C_1] = \sum_{i=1}^\infty \frac{n}{2^im^{(i-1)}} = \frac{mn}{2m-1}
\label{eqn:TournamentFirstPhase}
\end{equation}

During the second phase, we need to examine the balls in the lists as well as those in the discarded tuples. Let us first consider the tuples. The number of discarded tuples will depend on the number of comparisons during the first phase. A tuple is created whenever a comparison is performed and the two balls being compared have different colours. The expected number of comparisons during the first phase is given by equation (\ref{eqn:TournamentFirstPhase}) above, and the probability that two balls do not have the same colour is $1-1/m$. Heuristically, therefore, we can say that the expected number of discarded tuples is the product of equation (\ref{eqn:TournamentFirstPhase}) and $1-1/m$. 

But for each tuple we may have to perform one comparison (if the first ball in the tuple matches the candidate) or two comparisons (if the first ball in the tuple does not match the candidate). Since the probability of the first ball not matching the candidate is $1-1/m$, the expected number of comparisons per tuple is $2-1/m$. Putting these elements together, we can say that the expected number of comparisons in the second phase from the discarded tuples is:

\begin{equation}
E[C_{2D}] = \frac{mn}{2m-1}(1-\frac{1}{m})(2-\frac{1}{m}) = \frac{(1-2m)(1-m)n}{m(2m-1)}
\label{eqn:TournamentSecondPhaseDiscarded}
\end{equation}

Finally, we also need to consider the comparisons for the lists. Once the first phase of the algorithm is completed, the lists will be empty unless they initially had an odd number of balls in which case they will contain a single ball. The expected size of list $i$ is $n/(2m)^i$, but it is not possible to say whether this is odd or even in general. However, we can iterate through the lists from $i=0$ to $i=\log_2(n)$ and if the expected size is odd then add an additional comparison.

\subsection{Experimental Validation}

We validate our analysis with experiments in the same way as for the MJRTY algorithm.  The only difference is that for this algorithm we also include the case of two colours. The expected number of comparisons is found to always be within the 99\% confidence interval of the mean number of observed comparisons. The average relative error is 0.11\%, 0.08\%, 0.06\% and 0.05\% for 2, 3, 5 and 10 colours respectively.

\begin{figure}[t]
	\centering
	\includegraphics[width=0.9\columnwidth]{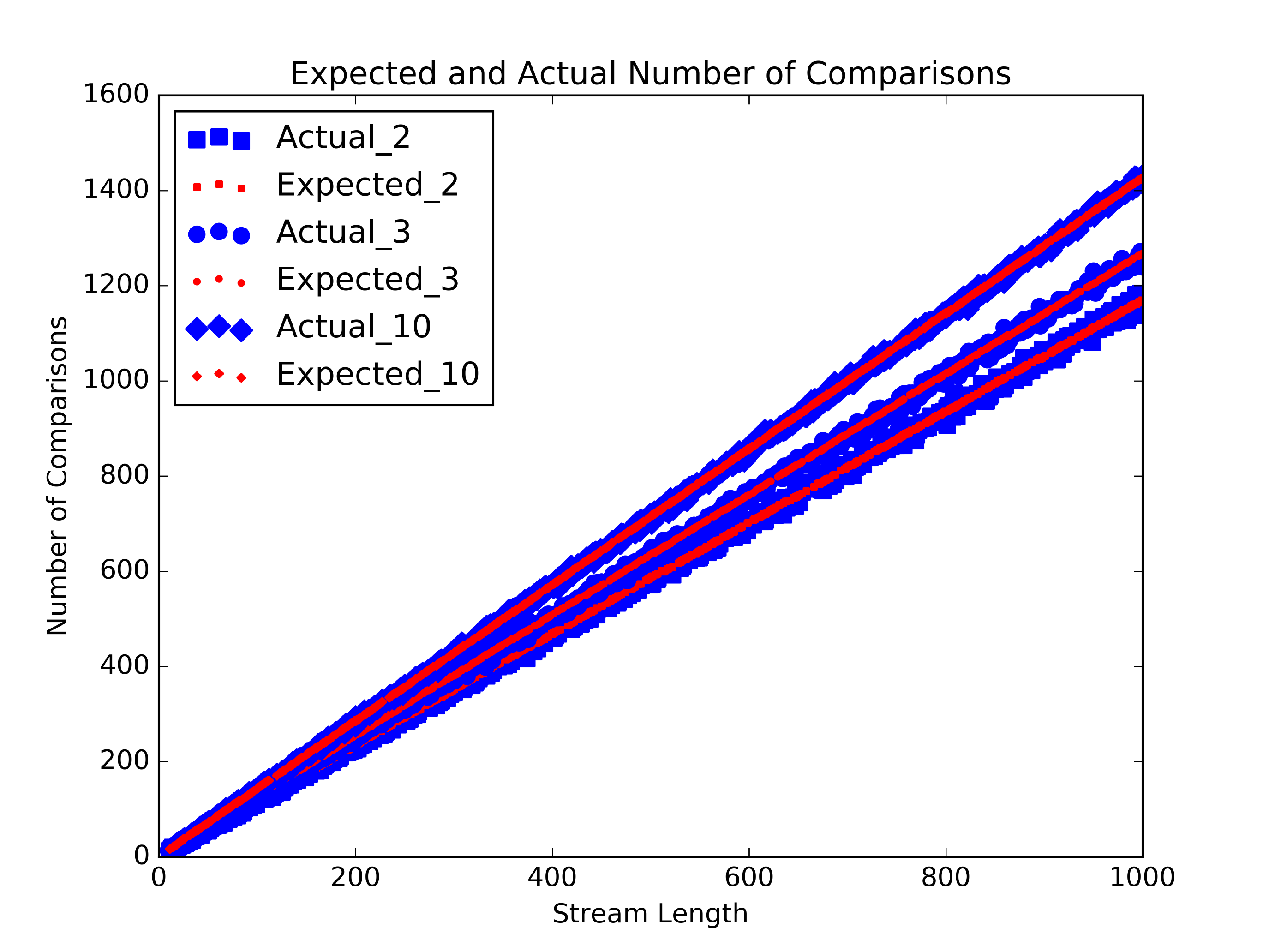}
	\caption{A comparison of the expected and actual number of comparisons for the Tournament algorithm for a number of different colours (best viewed in colour).}
	\label{fig:Tournament}
\end{figure}

Fig. \ref{fig:Tournament} shows the results of our second experiment with 2, 3 and 10 colours.

\section{Fischer and Salzberg's Algorithm}

\subsection{Brief Description of the Algorithm}

The Fischer-Salzberg algorithm is also based on the principle of pairing balls but with a slight twist. Rather than pairing and discarding, the algorithm attempts to create an ordering of paired balls by having a condition that no two adjacent balls can be of the same colour. We refer to this condition as the \textit{adjacency condition}. The idea is that, if there is a majority, it will be impossible to create such an ordering \footnote{There is one exception, in the case that $n$ is odd and the first, last and every second ball is of the same colour. We do not consider this case here explicitly for simplicity but the algorithm handles such a case with no additional comparison compared to a case of no majority}. 

The algorithm starts by creating a list which conforms to the adjacency condition. This is achieved through two data-structures: the \textit{list} and the \textit{bucket}. Balls are drawn from the stream and compared to the current last ball in the list. If the two balls have the same colour then they cannot be placed next to each other, so instead the drawn ball is placed in the bucket. If the two balls have different colours, the drawn ball is added to the end of the list and a ball from the bucket is added afterwards (if any exist). By taking balls from the bucket and appending them to the list, it is guaranteed that all balls in the bucket have the same colour, which is also the colour of the last ball in the list. It is also part of creating a list of all the balls without breaking the adjacency condition.

Once all the balls have been drawn, the last ball in the list is the candidate and will be of the majority, if one exists. However, balls may remain in the bucket and it may be possible to insert them into the list without violating the adjacency condition. If that is possible, then there is no majority. Therefore, a second phase is required which attempts to insert the balls from the bucket into the list without violating the adjacency condition.

During the second phase, the list is traversed in reverse order, always comparing the last ball in the list to the candidate. If the two are the same, then the last two balls in the list are discarded because balls from the bucket cannot be inserted either at the end of the list or between the last two balls (because all balls in the bucket will be the same colour as the candidate). If they are not the same, then the last ball in the list and one from the bucket are discarded because a ball from the bucket can be inserted at the end of the list without breaking the adjacency condition. The algorithm stops if a ball is ever needed from the bucket but the bucket is empty because then we have successfully found an ordering that does not break the adjacency condition and there is no majority colour. If the entire list is traversed and at least one ball remains in the bucket, then that ball is of the majority colour.

The pseudocode for the algorithm can be found in the Appendix.

\subsection{Analysis of the Algorithm}

The Fischer-Salzberg algorithm is extremely difficult to analyse because it is so sensitive to the order of the balls. Therefore, here we present a heuristic analysis that we show, empirically, to be extremely accurate.

The first phase of the algorithm will always use $n-1$ comparisons. The number of comparisons in the second phase depends on the size of the bucket and whether or not there is a majority. During the first phase, balls are added to the bucket if they are the same colour as the last ball in the list. Otherwise, a ball is removed from the bucket. The size of the bucket can therefore be modelled as a one-dimensional random walk with probability $1/m$ of going \textit{up} and probability $1-1/m$ of going \textit{down}. There is a barrier at 0 such that when the walk is at zero, \textit{down} means staying at 0.

Unfortunately, we were unable to solve this random walk to provide an expected position after $n$ steps. However, we can provide a heuristic analysis by treating the walk as if there was no barrier. When there is a majority, the probability of going up dominates. Let $\rho$ be the proportion of the majority coloured ball, then the expected position of the walk would be $n\rho - n(1-\rho) = n(2\rho-1)$ which would then be the expected size of the bucket after the first phase. The list would then have $n - n(2\rho-1) = 2n(1-\rho)$ balls in it. We would expect to traverse the list two balls at a time for $n(1-\rho)$ comparisons in the second phase.

To complete this part of the analysis we need to know the value of $\rho$ - the expected proportion of balls of the majority colour. $\rho$ is the expected value of a Bernoulli distribution with probability of success $1/m$ conditioned on there being a majority of successes. We found no closed form for this but it can be calculated directly from the definition of conditional expected value (taking into account that we do not care which colour is of the majority):

\begin{equation}
\rho = m\sum_{x=\frac{n}{2}+1}^{n} x {n\choose x}(1/m)^x(1-1/m)^{n-x}
\end{equation}

When there is no majority, the expected position of the walk is negative, equating to an empty bucket. The list length is therefore equal to $n$. We must traverse the list, two balls at a time, until we encounter a non-matching ball. When there are only two colours and no majority we will have to traverse the entire list for $n/2$ comparisons. When there are more than two colours, we can model the list as a binomial distribution with ``success'' defined as encountering a non-matching ball which happens with probability $1-1/m$. The expected number of trials before encountering the first success in a binomial distribution is the reciprocal of the probability of success so we would expect to have $1/(1-1/m) = m/(m-1)$ comparisons in the second phase.

Finally, to produce an overall expected number of comparisons, we must compute the probability of there being a majority. Again we model our stream as a binomial distribution with probability of success $1/m$. The probability of having a minority of successes can be directly calculated from the cumulative distribution function and from that the probability of having a majority can be found. 

Let $P(m)$ be the probability of having a majority and let $E_m[C]$ be the expected number of comparisons when there is a majority. We can then say that the expected number of comparisons for the Fischer-Salzberg algorithm is:

\begin{equation}
E[C] = (n-1) + \left(P(m)E_m[C] + (1-P(m)m/(m-1)\right)
\label{eqn:FischerSalzbergExpected}
\end{equation}

\subsection{Experimental Validation}
We validate our analysis with experiments in the same way as for the MJRTY and Tournament algorithms.  The expected number of comparisons is found to always be within the 99\% confidence interval of the mean number of observed comparisons. The mean relative error was 0.03\%, 0.23\%, 0.04\% and 0.01\% for 2, 3, 5 and 10 colours respectively. The relatively high mean relative error with 3 colours is because at small stream sizes (250 and 500) the relative error was 2.27\% and 1.40\% which brought up the average.

\begin{figure}[t]
	\centering
	\includegraphics[width=0.9\columnwidth]{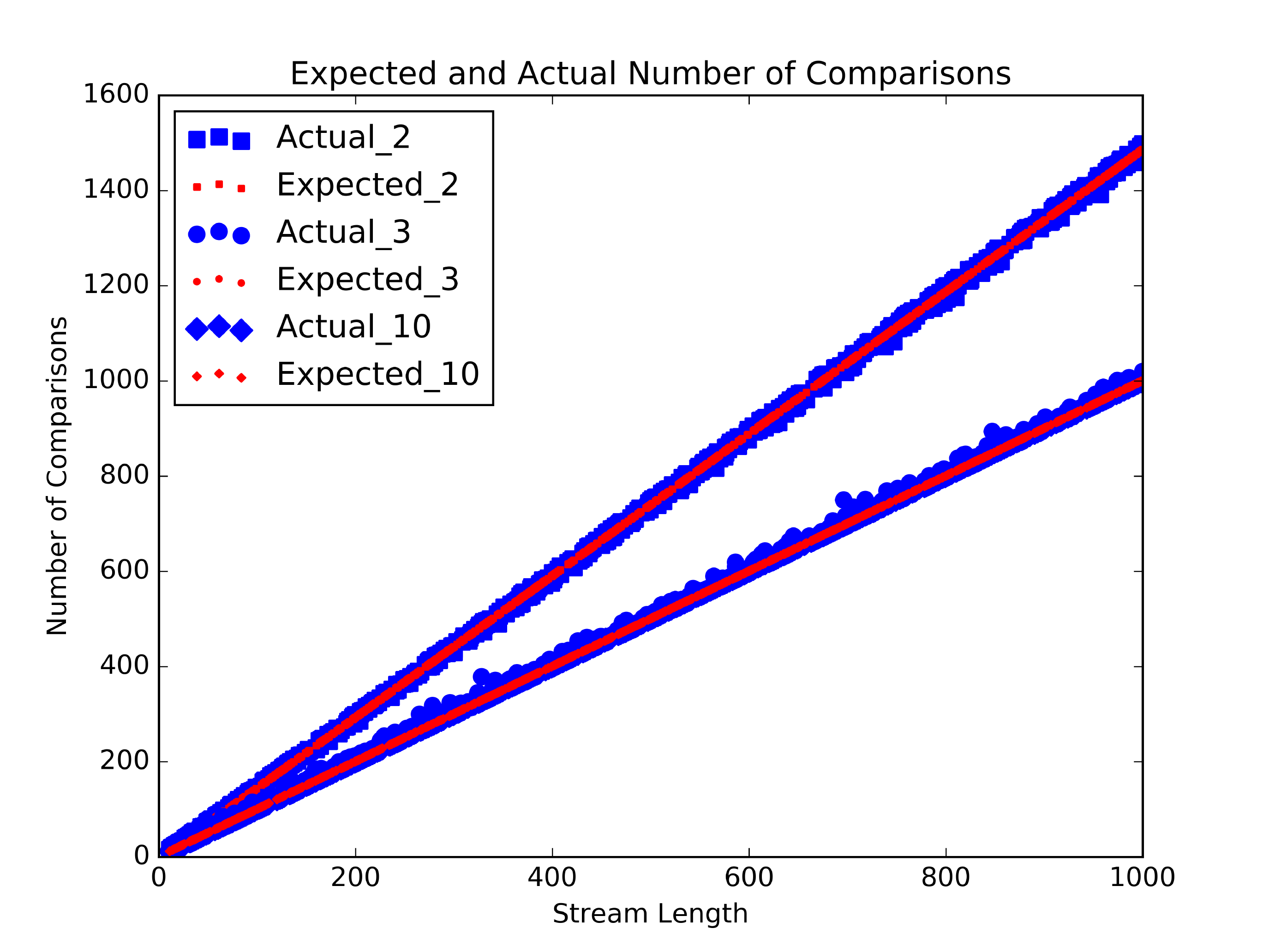}
	\caption{A comparison of the expected and actual number of comparisons for the Fischer-Salzberg algorithm for a number of different colours (best viewed in colour). The results for 3 and 10 colours overlap.}
	\label{fig:FischerSalzberg}
\end{figure}

Fig. \ref{fig:FischerSalzberg} shows the results of our second experiment. The number of comparisons with 3 and 10 colours are very similar and cannot be discerned in the graph.

\section{Conclusion}
\label{sec:Conclusion}

\begin{figure}[tp]
	\centering
	\includegraphics[width=\columnwidth]{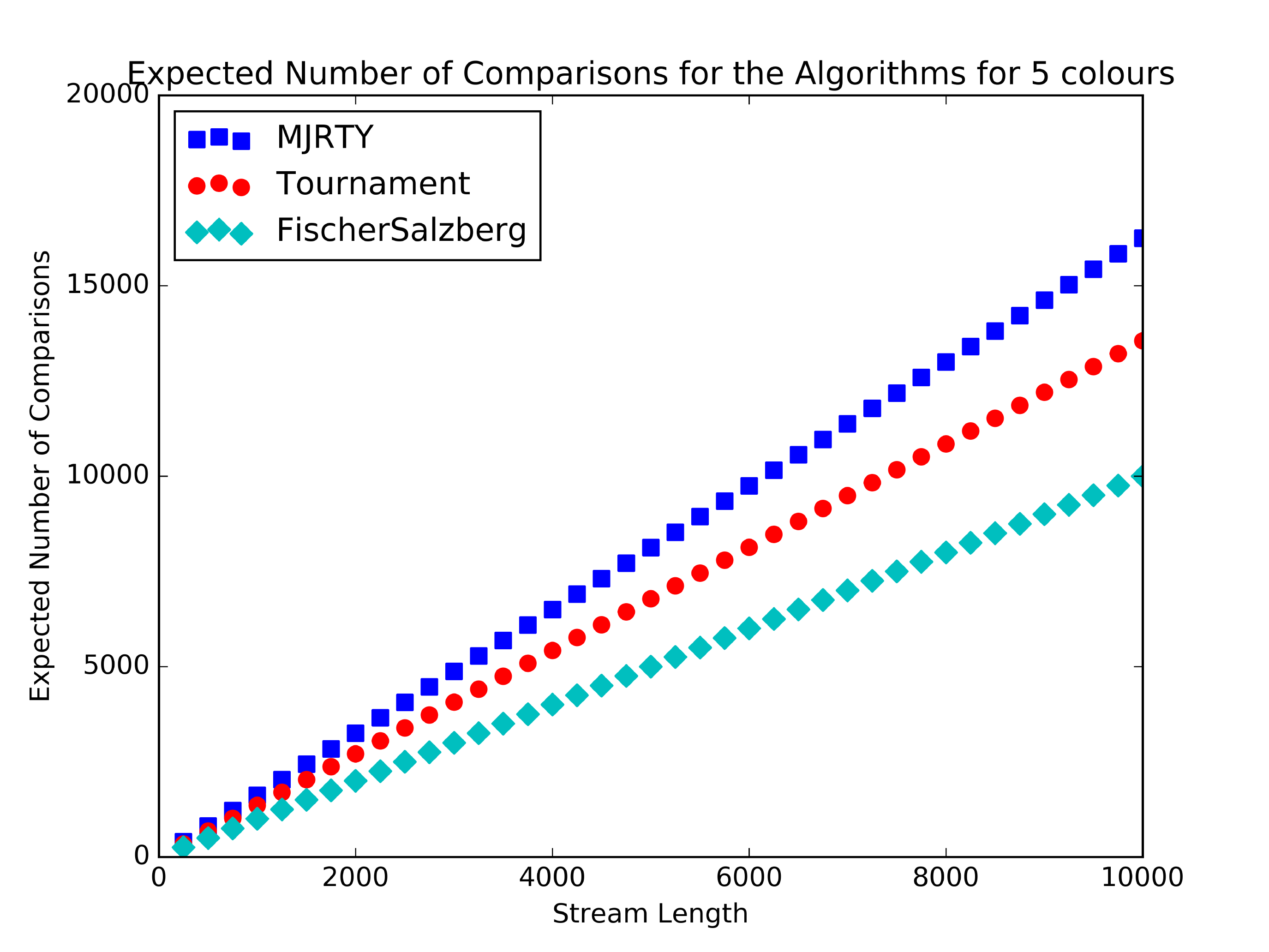}
	\caption{A comparison of the three algorithms with five colours and varying stream length.}
	\label{fig:AllLengths}
\end{figure}

\begin{figure}[tp]
	\centering
	\includegraphics[width=\columnwidth]{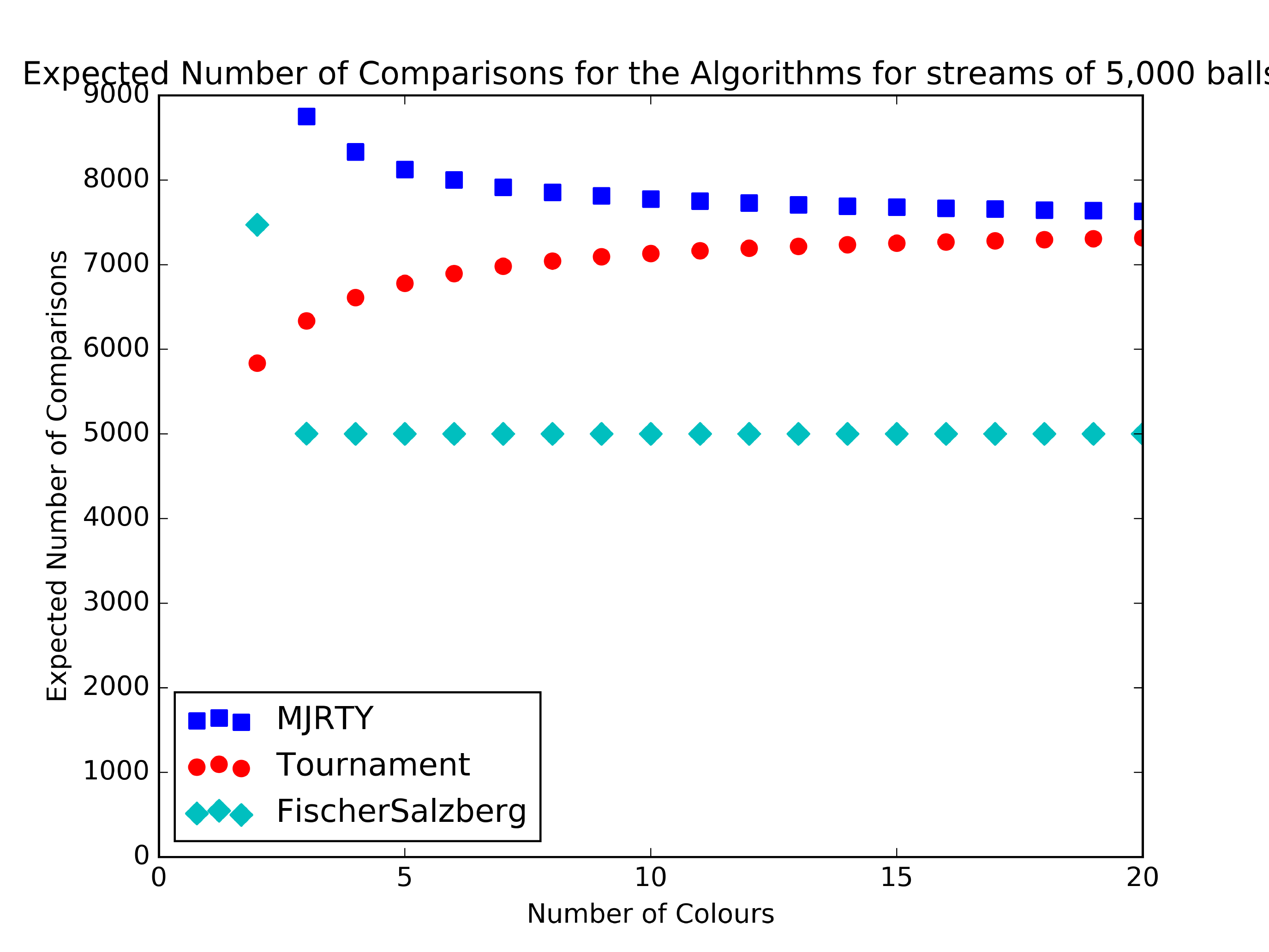}
	\caption{A comparison of the three algorithms with varying number of colours for a stream of 5,000 balls.}
	\label{fig:AllColours}
\end{figure}

The majority problem is an interesting problem that is a special case of the heavy hitters problem. In this paper we have presented a heuristic analysis of the expected number of comparisons required by the three deterministic algorithms that appear in the literature, assuming an arbitrary number of colours. 

Fig. \ref{fig:AllLengths} shows a direct comparison of the three algorithms when varying the stream length but keeping the number of colours constant at five colours. Fig. \ref{fig:AllColours} shows a direct comparison of the three algorithms with a constant stream size of 5,000 balls but varying the number of colours.

It is interesting to note that the most-widely discussed of the algorithms (Boyer and Moore's MJRTY) algorithm requires the largest number of comparisons. In contrast, the Fischer-Salzberg algorithm, which has been described as being ``essentially identical'' \cite{Cormode2010}, requires the fewest. Both algorithms have the same number of comparisons in the first phase but Fischer-Salzberg does some extra work in the first phase to save comparisons in the second phase. 

The Tournament algorithm also requires fewer comparisons than MJRTY. This is in both the first and second phase. The trade-off is the overhead of maintaining various lists.

These results suggest that measuring the number of comparisons is not sufficient to fully characterise the performance of these algorithms. A fuller investigation is warranted which considers the actual execution times.

Nevertheless, our analysis shows that the Tournament algorithm has the fewest expected comparisons in the first phase. It has an expected number of $mn/(2m-1)$ which, as the number of colours increases, tends to $n/2$. This compares to $n-1$ for both of the other algorithms. Since much of the later work on the more general heavy hitters problem does not consider verification (i.e. the second phase), this lower number of expected comparisons suggests that variations of the Tournament algorithm may outperform variants of the MJRTY algorithm.

Furthermore, there have been proposals for a parallel version of the MJRTY algorithm \cite{Lei1993,Cafaro2011}. The Tournament algorithm, however, seems better suited for a parallel implementation and it would be interesting to see whether the overheads of parallelism result in the Tournament algorithm being more efficient than MJRTY.

\bibliographystyle{elsart-num}
\bibliography{AverageCaseMajority}

\newpage

\appendix

\section{Pseudocode for the MJRTY Algorithm}
\label{App:AppendixA}

\begin{algorithm}[h!]
	\caption{Pseudocode for the MJRTY Algorithm}
	\label{alg:MJRTY}
	\begin{algorithmic}[1]
		\STATE c $\gets 0$
		\FOR{$i \gets 1$ to $n$}
		\IF{$c = 0$}
		\STATE $j \gets i$
		\STATE $c \gets 1$
		\ELSIF{$x_i = x_j$}
		\STATE $c \gets c + 1$
		\ELSE
		\STATE $c \gets c - 1$
		\ENDIF
		\ENDFOR
		\IF {$c = 0$}
		\STATE No majority
		\ELSE
		\STATE $c \gets 0$
		\FOR{$i \gets 1$ to $n$}
		\IF{$x_i = x_j$}
		\STATE $c \gets c+1$
		\IF{$c > n/2$}
		\STATE Majority is of colour $x_j$
		\ENDIF
		\ENDIF
		\ENDFOR
		\STATE No majority
		\ENDIF
		\end{algorithmic}
		\end{algorithm}
		

\newpage

\section{Pseudocode for the Fischer-Salzberg Algorithm} \label{App:AppendixB}		

\begin{algorithm}[h!]
	\caption{Pseudocode for the Fischer-Salzberg Algorithm}
	\label{alg:FS}
	\begin{algorithmic}[1]
		\STATE $l=0$
		\FOR{$ i \gets 1$ to $n$}
		\IF{$x_i = $ list[$l$]}
		\STATE bucket.append($x_i$)
		\ELSE
		\STATE list.append($x_i$)
		\STATE $l$++
		\IF{bucket.empty()}
		\STATE list.append(bucket.pop())
		\STATE $l$++
		\ENDIF
		\ENDIF
		\ENDFOR
		\STATE $C=$list[$l$]
		\WHILE{$!$list.empty()}
		\IF{list.pop()=$C$}
		\STATE list.pop()
		\ELSE
		\IF{bucket.empty()}
		\STATE No majority
		\ELSE
		\STATE bucket.pop()
		\ENDIF
		\ENDIF
		\ENDWHILE
		\STATE Majority is $C$
	\end{algorithmic}
\end{algorithm}

\newpage

\section{Pseudocode for the Tournament Algorithm} \label{App:AppendixC}	

\begin{algorithm}[h!]
	\caption{Pseudocode for the Tournament Algorithm}
	\label{alg:Tournament}
	\begin{algorithmic}[1]
		\STATE $i \gets 0$
		\WHILE{! list[$i$].empty()}
		\STATE $j \gets 0$
		\WHILE{$j < $ list[$i$].length() \&\& list[$i$].length() $> 1$}
		\IF{list[$i$][$j$] == list[$i$][$j+1$]}
		\STATE list[$i+1$].append(list[$i$][$j$])
		\ELSE
		\STATE tuples.add(list[$i$][$j$], list[$i$][$j+1$], $i$)
		\ENDIF
		\STATE $j \gets j+2$
		\ENDWHILE
		\STATE $i++$
		\ENDWHILE
		\WHILE{$i > 0$}
		\IF{list[$i$].length() $== 1$}
		\IF{Candidate == NULL}
		\STATE Candidate = list[$i$][$0$]
		\STATE $c = 2^i$
		\ELSE
		\IF{list[$i$][$0$] == Candidate}
		\STATE$c \gets c + 2^i$
		\ENDIF
		\ENDIF
		\ENDIF
		\STATE $i--$
		\ENDWHILE
		\FORALL{tuples}
		\IF{tuple[$0$] == Candidate OR tuple[$1$] == Candidate}
		\STATE $c \gets c + 2^{tuple[2]}$
		\ENDIF
		\ENDFOR
		\IF{$c > n/2$}
		\STATE Candidate is majority
		\ELSE
		\STATE No majority
		\ENDIF
	\end{algorithmic}
\end{algorithm}

\end{document}